\documentclass{emulateapj}
\usepackage{float}
\usepackage{amsmath}

\def\app#1#2{%
  \mathrel{%
    \setbox0=\hbox{$#1\sim$}%
    \setbox2=\hbox{%
      \rlap{\hbox{$#1\propto$}}%
      \lower1.1\ht0\box0%
    }%
    \raise0.25\ht2\box2%
  }%
}

\begin{document}

\title{Velocity Dispersion, Size, S\'ersic Index and $D_n4000$: The Scaling of Stellar Mass with Dynamical Mass for Quiescent Galaxies}

\author{H. Jabran Zahid \& Margaret J. Geller} 
\affil{Smithsonian Astrophysical Observatory, Harvard-Smithsonian Center for Astrophysics - 60 Garden Street, Cambridge, MA 02138}

\def\mean#1{\left< #1 \right>}

\begin{abstract}

We examine the relation between stellar mass, velocity dispersion, size, S\'ersic index and $D_n4000$ for ~40,000 quiescent galaxies in the SDSS. At a fixed stellar mass, galaxies with higher $D_n4000$ have larger velocity dispersions and smaller sizes. $D_n4000$ is a proxy for stellar population age, thus these trends suggest that older galaxies typically have larger velocity dispersions and smaller sizes. We combine velocity dispersion and size into a dynamical mass estimator, $\sigma^2 R$. At a fixed stellar mass, $\sigma^2 R$ depends on $D_n4000$. The S\'ersic index is also correlated with $D_n4000$. The dependence of $\sigma^2 R$ and S\'ersic index on $D_n4000$ suggests that quiescent galaxies are not structurally homologous systems. We derive an empirical correction for non-homology which is consistent with the analytical correction derived from the virial theorem. After accounting for non-homologous galactic structure, we measure $M_\ast \propto M_d^{0.998 \pm 0.004}$ where $M_\ast$ is the stellar mass and $M_d$ is the dynamical mass derived from the velocity dispersion and size; stellar mass is directly proportional to dynamical mass. Quiescent galaxies appear to be in approximate virial equilibrium and deviations of the fundamental plane parameters from the expected virial relation may result from mass-to-light ratio variations, selection effects and the non-homology of quiescent galaxies. We infer the redshift evolution of velocity dispersion and size for galaxies in our sample assuming purely passive evolution. The inferred evolution is inconsistent with direct measurements at higher redshifts. Thus quiescent galaxies do not passively evolve. Quiescent galaxies have properties consistent with standard galaxy formation in $\Lambda$CDM. They form at different epochs and evolve modestly increasing their size, velocity dispersion and S\'ersic index after they cease star formation.

\end{abstract}

\section{Introduction}

Observable properties of galaxies such as luminosity, stellar mass, morphology, size, stellar velocity dispersion and/or rotational velocity, gas mass, star formation rate, metallicity and stellar population parameters like color and/or age are correlated with one another \citep{Faber1976, Kormendy1977, Tully1977, Lequeux1979, Haynes1984, Roberts1994, Gavazzi1996b, Kennicutt1998a, Brinchmann2000, Kauffmann2003a, Noeske2007a}. Although galaxies appear to be complex and diverse, the interdependence of these properties indicates that only a subset of the parameters is fundamental to understanding galaxy evolution; identifying these parameters is not trivial \citep{Disney2008}. A robust theory of galaxy evolution requires identification of the fundamental properties and the relevant physical processes and initial conditions affecting these properties.


The age of the stellar population reflects the formation epoch of the galaxy. More massive quiescent galaxies in the local universe tend to have older stellar populations \citep[e.g.,][]{Kauffmann2003a, Gallazzi2005, Thomas2005}. These stellar population age trends can be measured directly from spectral indicators. Quantifying the dependence of galaxy properties on stellar population age provides a means to study galactic evolution from a sample of galaxies at a single epoch. This so-called ``archeological" approach is an alternative to measuring galaxy population properties at different epochs and then inferring evolution. Reconciliation of the inferred evolution of fundamental galaxy properties based on these approaches is an important test of self-consistency.

Several studies suggest that the velocity dispersion is the best observable to connect galaxies to their dark matter halos \citep[]{Wake2012a, Wake2012b, Bogdan2015, Zahid2016c}. The scaling of the velocity dispersion with stellar mass \citep[i.e.,][]{Faber1976} connects these two fundamental properties of galaxies and the dependence of this relation on redshift provides important constraints on the evolution of galaxies. \citet{Zahid2016c} show that the relation between stellar mass and velocity dispersion does not evolve significantly with redshift for quiescent galaxies at $z<0.7$ \citep[see also][]{Shu2012, Montero-Dorta2016}. \citet{Belli2014} report mild evolution in the relation between stellar mass and velocity dispersion at  $0.9 < z< 1.7$,  \citep[see Figure 13 in][]{Zahid2016c}. They attribute this evolution to the smaller sizes of galaxies at high redshift compared to the local population \citep[see also][]{Belli2016}. This coevolution is expected if quiescent galaxies are in virial equilibrium.

Virial equilibrium of quiescent galaxies is supported by the scaling relation between velocity dispersion and stellar mass. \citet{Zahid2016c} find that $\sigma \propto M_\ast ^{0.3}$. This power-law index is consistent with virial equilibrium and is the same as the scaling between dark matter halo mass and dark matter halo velocity dispersion \citep{Evrard2008}. Thus, to first order, quiescent galaxies may be in virial equilibrium. If so, the central dynamical mass is either dominated by or proportional to the stellar mass \citep{Zahid2016c}. The virial scaling between stellar mass and velocity dispersion appears to persist to at least $z\sim1.7$ \citep{Belli2014}.

If quiescent galaxies indeed approximate virialized systems, their sizes and velocity dispersions are not independent. Thus the coevolution of these properties has important implications for the dynamical state of these galaxies. 

Sizes of galaxies appear to evolve with cosmic time. Quiescent galaxies at early times were significantly smaller on average than local quiescent galaxies at the same stellar mass \citep[e.g.][]{Daddi2005, Zirm2007, Buitrago2008, vanDokkum2008b, Damjanov2011, vanderwel2014}. These observations suggest that either individual quiescent galaxies grow with time, e.g., via mergers \citep{White2007, Naab2009, Bezanson2009, Newman2012}, and/or that quiescent galaxies added to the population at late times are larger \citep{vanDokkum2001, Carollo2013}. Given the dearth of velocity dispersion measurements, it is not clear whether the redshift coevolution of velocity dispersion and size is consistent with virial equilibrium.

Quiescent galaxies in the local universe show a tight relation between luminosity, velocity dispersion and size known as the fundamental plane \citep{Dressler1987, Djorgovski1987}. The fundamental plane reflects the correspondence between the potential and kinetic energy of the system vis-a-vis the virial theorem \citep{Binney2008}. However, the parameters defining the plane deviate from the virial parameters, i.e. the fundamental plane is ``tilted" \citep[e.g.,][]{Jorgensen1995, Bernardi2003c, Cappellari2006}. Moreover, despite the tightness of the fundamental plane, there is intrinsic scatter correlated with other galaxy properties \citep{Saglia1993, Jorgensen1995, Forbes1998, Terlevich2002, Gargiulo2009, Graves2009b}. The residual correlations provide clues for understanding the origin of tilt of the fundamental plane. The tilt is often attributed to stellar population effects, initial mass function variations, differential dark matter contribution and/or the fact that quiescent galaxies are not structurally homologous \citep[e.g.][]{Prugniel1996, Ciotti1996}. The tilt could also mean that galaxies deviate in some systematic way from virial equilibrium.

The dependence of the fundamental plane on redshift provides important constraints for the coevolution of fundamental galaxy properties. The fundamental plane may evolve with redshift \citep{vanDokkum1996,Kelson1997, Treu2005, Holden2010, Saglia2010, vandeSande2014, Zahid2015}. However, most (perhaps all) of this evolution may result from the evolution of the mass-to-light ratios of galaxies \citep[e.g.][]{vanDokkum1996, Zahid2015}. The effect of the mass-to-light ratio variations on the fundamental plane can be taken into account by deriving the stellar mass fundamental plane. \citet{Zahid2016a} show that stellar mass FP does not evolve significantly for galaxies at $z<0.6$. Mild evolution is reported at higher redshifts \citep{Bezanson2013}.



Here we investigate the virial properties of quiescent galaxies by examining the relation between stellar mass, velocity dispersion and size and its dependence on stellar population age inferred from the $D_n4000$ index. This approach avoids the covariance inherent in the fundamental plane due to the use of size and surface brightness and it explicitly accounts for mass-to-light ratio variations. In Section 2 we describe the data and we briefly review the virial theorem in Section 3. Our results are in Section 4. We demonstrate the importance of taking the stellar mass as the independent variable in Section 5. In Section 6 we discuss the results and conclude in Section 7. We adopt the standard cosmology $(H_0, \Omega_m, \Omega_\lambda)$ = (70 km s$^{-1}$, 0.3, 0.7) throughout. 

\section{Data, Methods and Sample Selection}

Here we present the sample (Section 2.1) and methods for determining the stellar mass (Section 2.2), the velocity dispersion (Section 2.3), the galaxy size and the brightness profile (Section 2.4) and the $D_n4000$ index (Section 2.5). We also describe the $K-$corrections we use to derive a volume limited sample of quiescent galaxies (Section 2.5). Figure \ref{fig:hist} shows a histogram of relevant galaxy properties of the final volume limited sample.

\subsection{Data}

We analyze the Main Galaxy Sample of $\sim900,000$ galaxies from the Sloan Digital Sky Survey (SDSS) DR12\footnote{http://www.sdss.org/dr12/} \citep{Alam2015}. The sample is magnitude limited to $r<17.8$ and covers $\sim10,000$ deg$^{2}$ in the redshift range $0\lesssim z \lesssim 0.3$ \citep{York2000}. The spectral range of the SDSS observations is $3800 - 9200 \mathrm{\AA}$ at a resolution of $R\sim1500$ at $5000\mathrm{\AA}$ \citep{Smee2013}. We use the $ugriz$ c-model magnitudes from the SDSS imaging data \citep{Stoughton2002, Doi2010}.

\subsection{Stellar Mass}

\begin{figure*}
\begin{center}
\includegraphics[width = 2 \columnwidth]{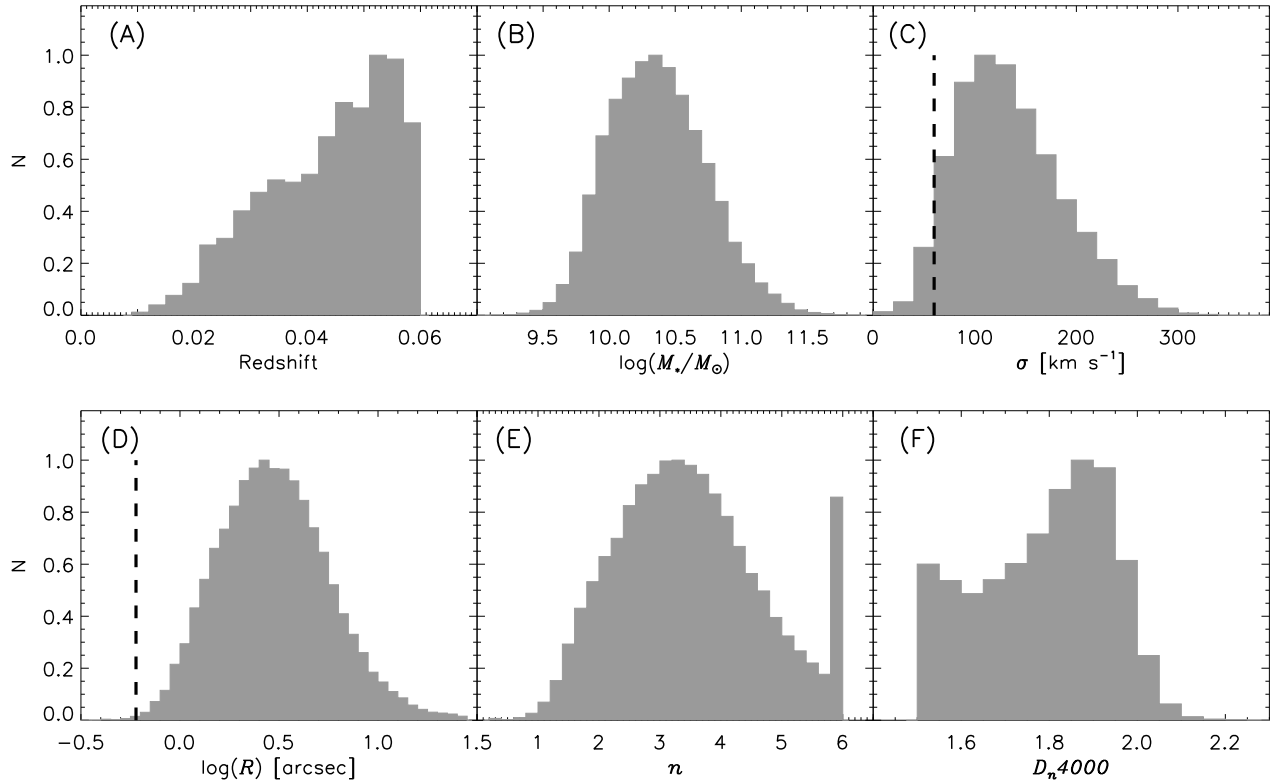}
\end{center}
\caption{(A) Redshift, (B) stellar mass, (C) velocity dispersion, (D) size, (E) S\'ersic index and (F) $D_n4000$ index distribution for the volume limited sample. The dashed lines in (C) and (E) show the resolution limits of the spectrograph and imaging data, respectively. The number of galaxies with velocity dispersions and sizes below the resolution limit is small. We do not remove these galaxies from the sample.}
\label{fig:hist}
\end{figure*}

We estimate the mass-to-light (M/L) ratio for each galaxy by $\chi^2$ fitting synthetic spectral energy distributions (SEDs) to the observed photometry. {The shape of the SED is used to derive the M/L ratio and we scale the measured luminosity by the M/L ratio to estimate the stellar mass.} We fit the observed photometry with {\sc Lephare} using the stellar population synthesis models of \citet{Bruzual2003} and the \citet{Chabrier2003} initial mass function (IMF). The model stellar populations have {three metallicities ($Z = 0.004$, 0.008 and 0.02)} and exponentially declining star formation histories (star formation rate $\propto e^{-t/\tau}$) with e-folding times of $\tau = 0.1,0.3,1,2,3,5,10,15$ and $30$ Gyr. We generate synthetic SEDs from these models by varying the extinction and stellar population age. We adopt the \citet{Calzetti2000} extinction law and allow $E(B-V)$ to range from 0 to 0.6. {The stellar population ages range between 0.01 and 13 Gyr. This procedure yields a distribution for the best-fit M/L ratio and stellar mass for each set of parameters. We adopt the median of this distribution.} Figure \ref{fig:hist}B shows the stellar mass distribution of the final volume limited sample. 

The SED fitting technique carries absolute uncertainties of $\sim0.3$ dex \citep{Conroy2009a}. Uncertainties arise from the parameters used to derive synthetic SEDs which include the star formation history, metallicity and dust extinction and also from the choice of stellar templates and IMF. By consistently estimating stellar masses for the whole sample, we achieve the relative accuracy required for this work; our analysis is not sensitive to absolute uncertainties. For reference, our stellar mass estimates are systematically smaller by 0.13 dex than the stellar mass estimates of the Portsmouth group (medianpdf in the Passive Kroupa stellar mass catalog). The slope of the relation between two mass estimates is unity.

\subsection{Velocity Dispersion}

Velocity dispersions are measured by the Portsmouth group from SDSS spectra \citep{Thomas2013}. The 3'' fiber apertures are centered on each galaxy thus the measurement is sensitive to the central region. \citet{Thomas2013} use the Penalized Pixel-Fitting (p{\sc PXF}) code \citep{Cappellari2004} with the \citet{Maraston2011} stellar population templates based on the MILES stellar library \citep{Sanchez-Blazquez2006}. The templates are matched to the instrument resolution and are parameterized by convolution with $\sigma$. The best-fit $\sigma$ is determined by minimizing the $\chi^2$ between the model and observed spectrum in the rest-frame wavelength range $4500-6500\mathrm{\AA}$. Throughout this work the velocity dispersion is quoted in km s$^{-1}$ and is referred to as $\sigma$. The typical velocity dispersion error for galaxies in the volume limited sample is $\sim5$ km s$^{-1}$.

The resolution of the SDSS spectrograph ($R\sim1500$) limits reliable estimates of the velocity dispersion to $\sigma>60$ km s$^{-1}$. Figure \ref{fig:hist}C shows the $\sigma$ distribution for the volume limited sample and the resolution limit; $5\%$ of the sample has $\sigma<60$ km s$^{-1}$. These galaxies are typically the least massive galaxies with the smallest $D_n4000$. Throughout this work, we calculate medians (not means) to characterize velocity dispersions as a function of various galaxy properties. Quantifying properties with the median mitigates bias due to the resolution limit. The $\sigma<60$ km s$^{-1}$ measurement are always smaller than the median of the population as a function of the galaxy properties we examine and thus do not bias the median. To minimize selection effects, we do not remove galaxies with small velocity dispersions from the sample.

\citet{Zahid2016c} compare velocity dispersion measurements from \citet{Thomas2013} with measurements for the same galaxies based on observations with Hectospec on MMT \citep{Fabricant2013}. The fiber aperture of Hectospec (1.5'') is half that of SDSS. From this comparison, \citet{Zahid2016c} derive an aperture correction 
\begin{equation}
\frac{\sigma_1}{\sigma_2} = \left(  \frac{R_1}{R_2}  \right)^{-0.033 \pm 0.011}.
\end{equation}
This correction is consistent with a correction derived from integral field spectroscopy \citep{Cappellari2006}. We correct all $\sigma$s to the measured half-light radius using this correction. After correcting for aperture effects and accounting for observational errors, \citet{Zahid2016c} find that the SDSS and Hectospec measurements are consistent \citep[see also][]{Fabricant2013}. The aperture correction is small; our results are insensitive to the correction.

{Quiescent galaxies may exhibit ordered rotation in addition to random stellar motions \citep[e.g.,][]{Emsellem2007}. Ordered rotation may contribute significantly to the velocity dispersion at large radii but has a negligible impact on the central stellar velocity dispersion \citep{Riciputi2005}.}

\subsection{Galaxy Size and Profile}

Sizes and profiles of SDSS galaxies are measured by the NYU group \citep{Blanton2005a, Blanton2005b, Padmanabhan2008} by fitting the photometry with a \citet{Sersic1968} model:
\begin{equation}
I(r) = A ~ \mathrm{exp} \left[-b(n) \left( [r/R]^{1/n} - 1 \right) \right].
\label{eq:sersic}
\end{equation}
Here, $r$ is the radial coordinate and $I(r)$ is the seeing-corrected radial galaxy profile. The free parameters of the model are the half-light radius $R$ and the S\'ersic index $n$. $n$ ranges between 0 and 5.9 with $n=1$ and $n=4$ yielding the exponential and de Vaucouleurs profiles, respectively. {$A$ is the intensity at $R$ and $b(n)$ is defined such that half the galaxy light is contained within $R$.}

\citet{Blanton2005b} fit the S\'ersic model to the 1D radial profile $I(r)$ measured by taking the mean flux in annuli centered on the galaxy profile peak. To account for the seeing, the S\'ersic model is convolved with a Gaussian seeing model prior to fitting. The typical seeing is $\gtrsim1''.2$ \citep{Stoughton2002}, thus measurements of the radius are limited to half this value. We adopt the $r$-band $R$ and $n$ measured by the NYU group as measures of size and S\'ersic index, respectively. {Among various size estimates available \citep[e.g. SDSS pipeline;][]{Simard2011} the NYU group size estimates are in the best agreement with high-quality measurements from HST. The NYU group sizes are systematically larger than HST sizes by $\sim0.07$ dex but show no systematic trends \citep[see Figure 1 in][]{Zahid2016a}. We also analyze the $g$-band parameters and find that our results and conclusions are robust to the choice of photometric band.} Figures \ref{fig:hist}D and \ref{fig:hist}E show the size and S\'ersic index for the volume limited sample, respectively.

The seeing-limited resolution of the SDSS imaging sets the minimum size to $\gtrsim0''\!\!.6$. Figure \ref{fig:hist}E shows the size distribution and the resolution limit. Our volume limited selection ensures that nearly all the galaxies in our sample have sizes that are significantly larger than the resolution limit. Thus, imaging resolution is not a significant source of bias. 

Systematic errors related to the sky level used in the measurements does bias the measurements of $R$ and $n$ \citep{Blanton2005b}. For large S\'ersic index objects the S\'ersic index is underestimated \citep{Blanton2005b} and the bias appears to scale with the measurement \citep[see Appendix A of][]{Guo2009}. However, \citet[see appendix]{Taylor2010b} show that the covariance between the S\'ersic index and the size mitigates the systematic errors in dynamical masses measured from these quantities. 

The S\'ersic index returned by the \citet{Blanton2005b} procedure is limited to $n_s < 5.9$. This limit produces the spike at $n_s = 5.9$ in Figure \ref{fig:hist}E. This limit affects 5\% of the sample. Because of our binning procedure, the results are insensitive to this artificial truncation of the S\'ersic index.

\subsection{$D_n4000$ Index}

\begin{figure*}
\begin{center}
\includegraphics[width = 2 \columnwidth]{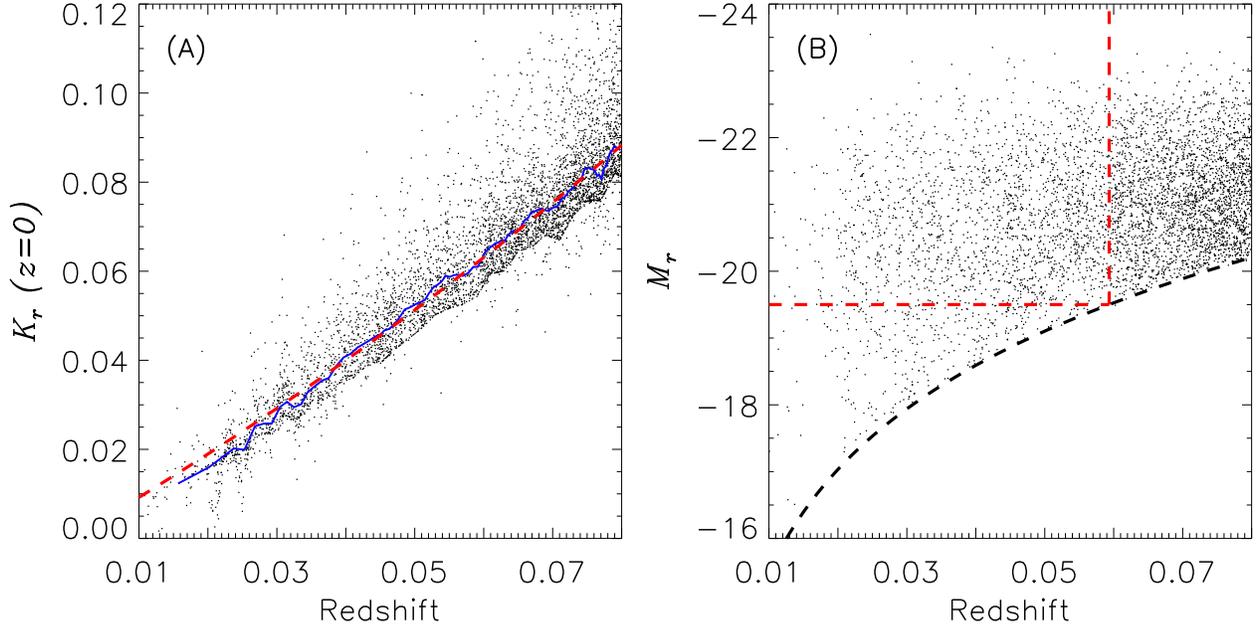}
\end{center}
\caption{(A) $r-$band $K-$correction for the sample as a function of redshift. Black points are individual galaxies. For clarity, we plot only a random sub-sample of the parent sample. The blue curve is the median $K-$correction in equally populated bins of redshift and the red curve is a second degree polynomial fit to the median. (B) $K-$corrected absolute $r-$band magnitude, $M_r$, as a function of redshift. The black dashed curve shows the limiting magnitude of $m_r<17.77$. Red lines indicate the $M_r$ and redshift limits of the volume limited sample. }
\label{fig:kcorr}
\end{figure*}

{The $D_n4000$ index is the ratio of the flux in two spectral windows adjacent to the 4000$\mathrm{\AA}$ break \citep{Balogh1999}: $3850-3950\mathrm{\AA}$ and $4000-4100\mathrm{\AA}$. We adopt the $D_n4000$ values reported in the MPA/JHU catalog \citep{Kauffmann2003a}.}

The $D_n4000$ index increases as function of stellar population age \citep{Kauffmann2003a, Zahid2015}. Thus, it is a directly measured spectral feature that is a proxy for the age of the stellar population. The index distribution is bimodal \citep{Kauffmann2003a, Geller2014} and we use it to select quiescent galaxies containing older stellar populations where the stellar kinematics are typically dominated by random motions \citep[e.g.,][]{Brinchmann2004}. The bimodality of the distribution occurs because the stellar populations of quiescent galaxies are dominated by older stars \citep[e.g.,][]{Kauffmann2003a}. The MPA/JHU\footnote{http://wwwmpa.mpa-garching.mpg.de/SDSS/DR7/} group measure the index for the SDSS Main Galaxy Sample and we adopt their measurements \citep{Kauffmann2003a}.

The $D_n4000$ index increases monotonically with time for stellar populations undergoing passive evolution. We quantify the time evolution of the $D_n4000$ index using the Flexible Stellar Population Synthesis (FSPS) model \citep{Conroy2009a, Conroy2010}. FSPS produces a model for any arbitrary star formation and metallicity history. For simplicity, we adopt a constant star formation rate for 1 Gyr and solar metallicity for the star formation and metallicity history of the model galaxy, respectively \citep[see][]{Zahid2015}. We calculate the $D_n4000$ index directly from the model spectra using the same procedure applied to the observations. We use this model result in Section 6.2 to link galaxies in our sample to quiescent galaxies at higher redshifts.

For stellar populations with $D_n4000>1.5$ the index is insensitive to the star formation history but does depend on metallicity. For quiescent galaxies, measurements indicate that at a fixed stellar mass the heavy element abundance evolution with redshift appears to be negligible \citep{Choi2014}. We examine the relative change in the average galaxy properties and their dependence on $D_n4000$ at a fixed stellar mass. Thus, our results are insensitive to metallicity variations.

Stellar population properties may vary spatially in a galaxy. A fixed fiber aperture size corresponds to varying physical size depending on galaxy distance. Thus, strong $D_n4000$ gradients in the centers of galaxies could introduce systematic effects in the measurement. \citet{Fabricant2008} compare $D_n4000$ index measurements from spectra observed through 1.5 and 3'' Hectospec and SDSS fiber apertures, respectively. They find that the $D_n4000$ index measured for the same galaxy varies systematically by $<5\%$. Thus, $D_n4000$ index gradients in the central regions of galaxies are small and do not significantly impact our results.

There is no single objective method to select quiescent galaxies \citep[e.g.,][]{Moresco2013}. We classify quiescent galaxies using the $D_n4000$ spectral indicator. Figure \ref{fig:hist}F shows that the $D_n4000$ distribution is bimodal and that our selection limit of $D_n4000 > 1.5$ is smaller than the minimum of the bimodal separation near $D_n4000\sim1.6$. Thus, our sample includes galaxies where the stellar kinematics may be dominated by ordered rotation rather than random motions. However, our approach is to examine the dynamical properties of galaxies as a function of $D_n4000$. This approach provides a means to identify systematic trends that may be related to contamination and/or sample selection. Because we analyze the sample as a function of $D_n4000$, the results are insensitive to the exact value of the $D_n4000$ selection threshold.

{$D_n4000$ is measured using data ranging between $3850 - 4100\mathrm{\AA}$ whereas velocity dispersion is measured from spectra at wavelengths $>4500\mathrm{\AA}$. The non-overlapping spectral ranges mitigate covariance between the velocity dispersion and the $D_n4000$ index.}

\subsection{The Volume Limited Sample Selection}

We select a volume limited sample of quiescent galaxies from our parent sample with the following properties:
\begin{itemize}
\item $m_r < 17.77$
\item $D_n4000 > 1.5$
\item $z > 0.01$
\item $0 < R < 27''\!\!.9 $.
\end{itemize}
The magnitude cut limits galaxies to the magnitude limited sample where the completeness is $\sim95\%$ \citep{Strauss2002}. The lower limit on $D_n4000$ selects quiescent galaxies. The lower redshift limit minimizes distance uncertainties produced by peculiar motions. The lower size limit ensures that a size and S\'ersic index are fit for each galaxy in the sample and the upper limit is set by the measurement procedure \citep{Blanton2005b}. The size selection effects $<1\%$ of the sample. These selection criteria yield a parent sample of 404,000 galaxies.

We select a volume limited sub-sample of galaxies from the parent sample by $K-$correcting the $r-$band magnitude to $z = 0$ using the correction given in the Value Added Galaxy Catalog of the NYU group \citep{Blanton2005a}. {The root-mean-square difference between the $K-$correction calculated by the NYU group and the one calculated using LePHARE is 0.01 magnitudes; the results are insensitive to the $K-$correction prescription.}  $K-$correction is necessary to account for the redshifting of the photometric bands. Figure \ref{fig:kcorr}A shows the $r-$band $K-$correction as a function of redshift. The red curve is a polynomial fit to the median $K-$correction as a function of redshift
\begin{equation}
K_r(z) = 0.90z + 2.59 z^2.
\end{equation}
We apply a $K-$correction to each galaxy and we apply the median correction to the magnitude limit. Figure \ref{fig:kcorr}B shows the volume limited selection for $M_r < -19.5$. The final volumed limited sample consists of $37,874$ galaxies. The volume limit selects galaxies with stellar masses $>10^{9.5} M_\odot$ and $z < 0.06$.

\section{Quiescent Galaxies and the Virial Theorem}

\begin{figure*}
\begin{center}
\includegraphics[width =  2\columnwidth]{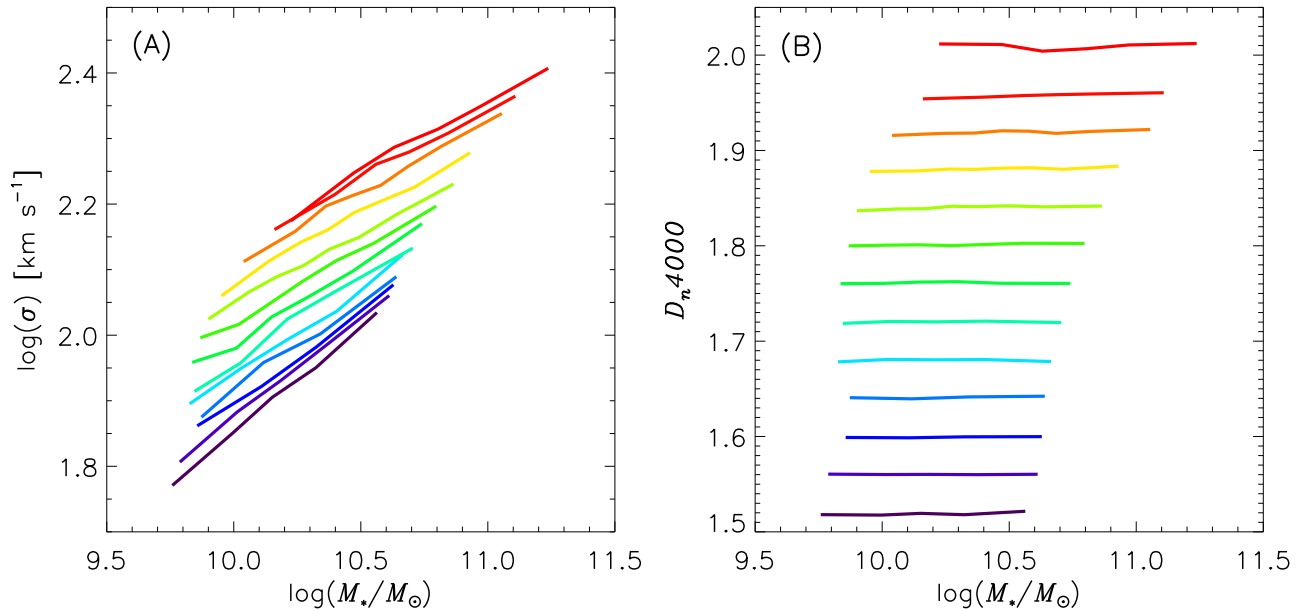}
\end{center}
\caption{(A) Median velocity dispersion as a function of stellar mass and $D_n4000$. (B) Median $D_n4000$ values corresponding to the curves in (A). Color-coding indicates the different relations for the same data. The typical bootstrapped error for the median velocity dispersion in each bin is $0.006$ dex. For clarity, error bars are omitted.}
\label{fig:vdisp_dn4000}
\end{figure*}

Here we review the application of the virial theorem to quiescent galaxies. The scaler virial theorem is
\begin{equation}
U +   2T  = 0, 
\end{equation}
where $U$ is the potential energy and $T$ is the kinetic energy of the system. The potential energy of a self-gravitating system is the gravitational binding energy
\begin{equation}
U=-\frac{k_U GM_d^2}{r}
\end{equation}
where $G$ is the gravitational constant, $M_d$ is the total dynamical, i.e. gravitational,  mass enclosed within a radius $r$ and $k_U$ depends on the mass distribution (e.g., for a uniform density sphere $k_U=3/5$). The kinetic energy of the system is
\begin{equation}
T = \frac{1}{2} M_d V^2 = k_T M_d \sigma^2,
\end{equation}
where $V^2$ is the mean-squared stellar velocity of the system. $k_T$ is a constant relating velocity dispersion $\sigma$ to $V$ which depends on the mass and velocity distribution of the system \citep[e.g.,][]{Ciotti1991}. Combining the three equations above we have
\begin{equation}
M_d = K_d \frac{\sigma^2r_e}{G},
\label{eq:md}
\end{equation}
where the constant $K_d$ combines the dependence of the kinetic and potential energies on the mass and velocity distribution of the system. Given an arbitrary projected density profile, $K_d$ can be computed via numerical integration. Solutions for the S\'ersic family of profiles have been explored in detail \citep[e.g.][]{Ciotti1991, Ciotti1997, Prugniel1997, Graham1997, Bertin2002, Mazure2002}. \citet{Bertin2002} provide the fitting formula for spherical, isotropic, non-rotating galaxies {accounting for projection effects}
\begin{equation}
K_d(n) = \frac{73.32}{10.465 + (n-0.94)^2} + 0.954.
\label{eq:k}
\end{equation}
Here, $n$ is the S\'ersic index (see Equation \ref{eq:sersic}). The relation has a numerical accuracy of $\sim1\%$ for $1<n<10$.

Equation \ref{eq:md} relates the dynamical mass, velocity dispersion and size of a galaxy in virial equilibrium. At a fixed dynamical mass, galaxies with larger sizes have smaller velocity dispersions and vice versa. If the dynamical mass of quiescent galaxies is proportional to the stellar mass, then an anti-correlation between $\sigma$ and $R$ is expected at a fixed stellar mass.

The central mass of a quiescent galaxy is comprised primarily of stars and dark matter. Thus, the dynamical mass in Equation \ref{eq:md} is the total (dark matter + stellar) gravitational mass. We are unable to constrain the dark matter contribution to the dynamical mass with the data because the fractional dark matter contribution is degenerate with absolute uncertainties in the stellar mass. In light of this degeneracy, our analysis depends only on the \emph{relative} accuracy of the stellar mass measurements and we do not draw any conclusions regarding the magnitude of the dark matter contribution.


\section{Velocity Dispersion, Size and S\'ersic Index as a Function of Stellar Mass and $D_n4000$}

We examine galaxy properties as a function of stellar mass and $D_n4000$ by sorting galaxies into equally spaced bins of $D_n4000$. We adopt a $D_n4000$ bin width of 0.04 except for the largest $D_n4000$ bin where we take all galaxies with $D_n4000 > 1.98$. In each $D_n4000$ bin we calculate the median of various galaxy properties in equally populated bins of stellar mass. We select the stellar mass bin width to contain $\sim500$ galaxies. We calculate statistical uncertainties by bootstrapping the data in each bin.

Figure \ref{fig:vdisp_dn4000}A shows the median velocity dispersion as a function of stellar mass and $D_n4000$. The $D_n4000$ values corresponding to the curves in Figure \ref{fig:vdisp_dn4000}A are shown in Figure \ref{fig:vdisp_dn4000}B. At a fixed stellar mass, the velocity dispersion is strongly correlated with $D_n4000$ and the relations between stellar mass and velocity dispersion at different $D_n4000$s are nearly parallel. The velocity dispersion difference between the highest and lowest $D_n4000$ bin for galaxies with $M_\ast = 10^{10.5} M_\odot$ is 0.23 dex.

\begin{figure}
\begin{center}
\includegraphics[width =  \columnwidth]{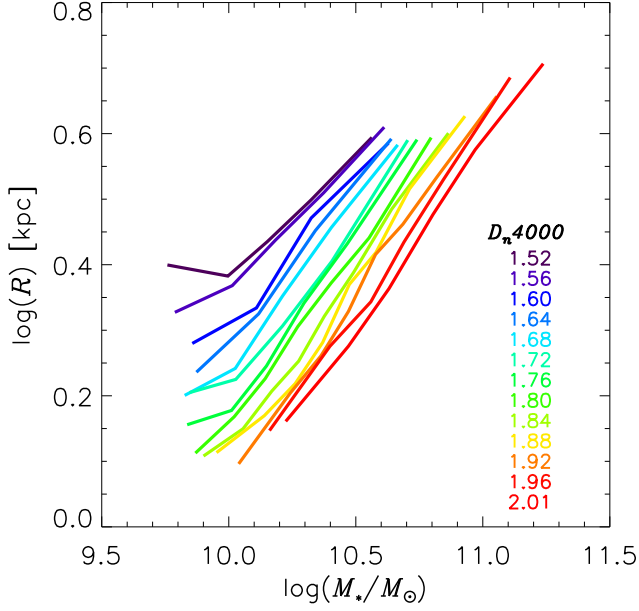}
\end{center}
\caption{Median size as a function of stellar mass and $D_n4000$. Median $D_n4000$ values corresponding to the curves are listed and are color-coded to match the curves plotted. These values are the same as the median $D_n4000$ plotted as a function of stellar mass in Figure \ref{fig:vdisp_dn4000}B. The typical bootstrapped error for the median size in each bin is $0.01$ dex. For clarity, error bars are omitted.}
\label{fig:size_dn4000}
\end{figure}

Figure \ref{fig:size_dn4000} shows the median size as a function of stellar mass and $D_n4000$. The trend between size and $D_n4000$ is opposite to the trend between velocity dispersion and $D_n4000$. At a fixed stellar mass, the size is \emph{anti}-correlated with $D_n4000$. Similar to the relation between velocity dispersion and stellar mass, the relations between size and stellar mass at different $D_n4000$s appear to be nearly parallel. The size difference between the highest and lowest $D_n4000$ bin for galaxies with $M_\ast = 10^{10.5} M_\odot$ is 0.26 dex.

Figures \ref{fig:vdisp_dn4000} and \ref{fig:size_dn4000} show that at a fixed stellar mass, galaxies with larger $D_n4000$ are smaller and have higher velocity dispersions. These trends are representative of the quiescent galaxy distribution in different $D_n4000$ bins (see Appendix) and are expected if galaxies are in approximate virial equilibrium and the stellar mass is proportional to the dynamical mass (see Equation \ref{eq:md}).

\begin{figure}
\begin{center}
\includegraphics[width =  \columnwidth]{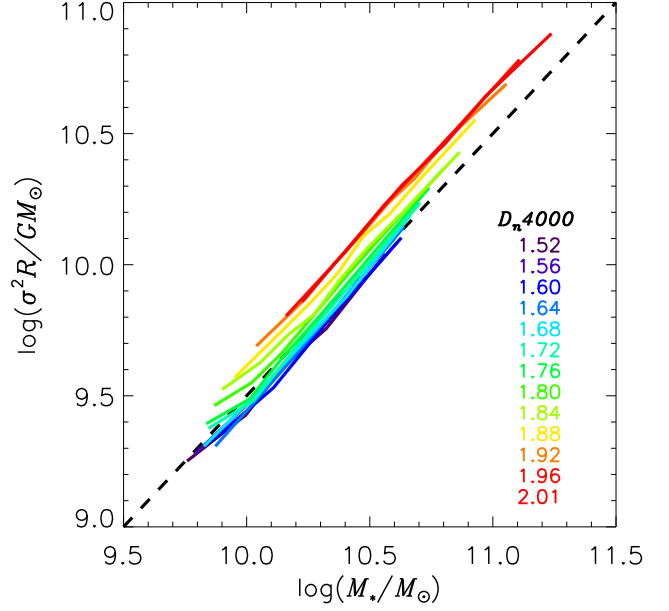}
\end{center}
\caption{Median $\sigma^2 R$ as a function of stellar mass and $D_n4000$. Median $D_n4000$ values corresponding to the curves are listed and are color-coded to match the curves plotted. These values are the same as the median $D_n4000$ plotted as a function of stellar mass in Figure \ref{fig:vdisp_dn4000}B. The dashed line is the one-to-one relation vertically offset by -0.5 dex.The typical bootstrapped error for $\sigma^2R$ in each bin is $0.01$ dex. For clarity, error bars are omitted.}
\label{fig:mdyn_dn4000}
\end{figure}

The virial theorem relates the dynamical mass to velocity dispersion and size; $M_d \propto \sigma^2 R$ (Equation \ref{eq:md}). Figure \ref{fig:mdyn_dn4000} shows $\sigma^2R$ as a function of stellar mass and $D_n4000$. At a fixed stellar mass, $\sigma^2R$ is correlated with $D_n4000$. The difference in $\sigma^2R$ between the highest and lowest $D_n4000$ bin for galaxies with $M_\ast = 10^{10.5} M_\odot$ is 0.20 dex. Given the scatter in $\sigma$ and $R$, $\sim0.7$ dex scatter is expected for $\sigma^2 R$ if the two quantities are uncorrelated. The measured scatter of $0.2$ dex is substantially smaller.

The dashed line in Figure \ref{fig:mdyn_dn4000} is the one-to-one relation offset by -0.5 dex in the vertical direction. Relations between stellar mass and $\sigma^2 R$ at different $D_n4000$s are parallel with a near unity slope. Thus, $\sigma^2 R \propto M_\ast$. This result implies that at a fixed $D_n4000$, the dark matter contribution in the centers of quiescent galaxies is a nearly constant and/or negligible fraction of the stellar mass. \citet{Zahid2016c} reach a similar conclusion on the basis of the virial scaling between the stellar mass and velocity dispersion of the quiescent galaxy population.

\begin{figure}
\begin{center}
\includegraphics[width = \columnwidth]{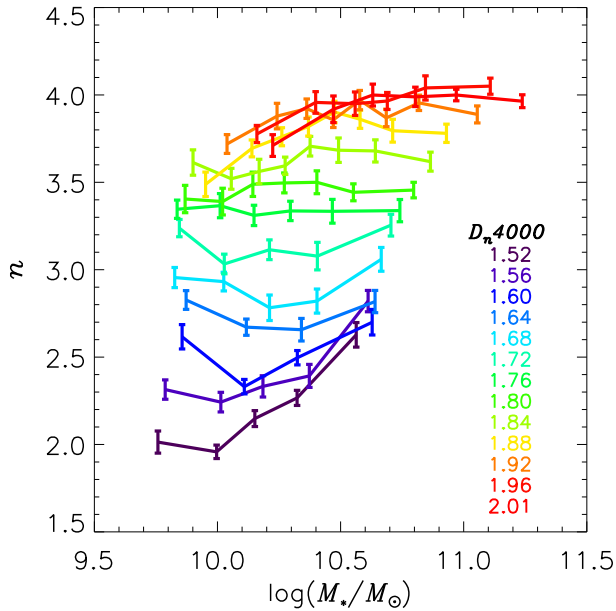}
\end{center}
\caption{Median S\'ersic index as a function of stellar mass and $D_n4000$. Median $D_n4000$ values corresponding to the curves are listed and are color-coded to match the curves plotted. These values are the same as the median $D_n4000$ plotted as a function of stellar mass in Figure \ref{fig:vdisp_dn4000}B. The error bars are the bootstrapped errors for the median $n$.}
\label{fig:sersic_dn4000}
\end{figure}

We investigate whether the correlation of $\sigma^2R$ and $D_n4000$ may be a consequence of variation in the structural properties of quiescent galaxies. Figure \ref{fig:sersic_dn4000} shows the relation between the S\'ersic index, stellar mass and $D_n4000$. At a fixed stellar mass, galaxies with large $D_n4000$ tend to have a larger S\'ersic index, i.e., steeper inner profiles and shallower outer profiles. \citet{Donofrio2011} report a similar trend between stellar mass, S\'ersic index and the stellar M/L ratio.

\begin{figure}
\begin{center}
\includegraphics[width =  \columnwidth]{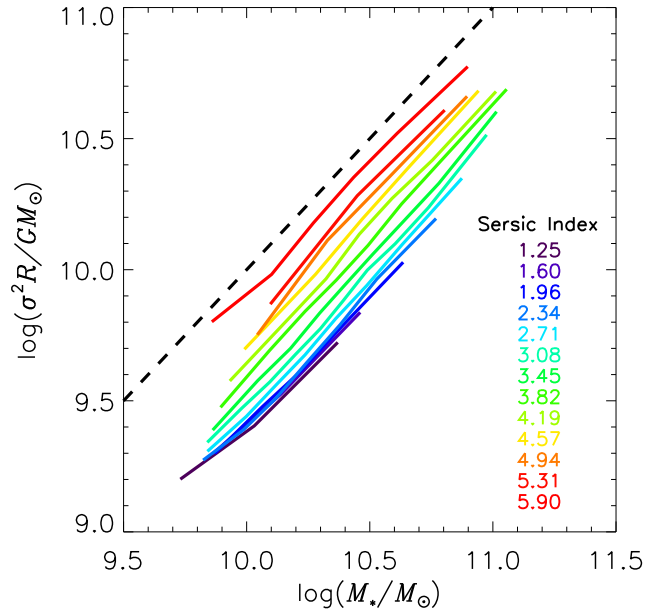}
\end{center}
\caption{{Median $\sigma^2 R$ as a function of stellar mass and S\'ersic index. Median S\'ersic index values corresponding to the curves are listed and are color-coded to match the curves plotted. The dashed line is the one-to-one relation.The typical bootstrapped error for $\sigma^2R$ in each bin is $0.01$ dex. For clarity, error bars are omitted.}}
\label{fig:mdyn_sersic}
\end{figure}

{Figures \ref{fig:mdyn_dn4000} and \ref{fig:sersic_dn4000} show that at a fixed stellar mass, $\sigma^2 R$ and S\'ersic index are both correlated with $D_n4000$. Thus, $\sigma^2 R$ and S\'ersic index are correlated. Figure \ref{fig:mdyn_sersic} shows the relation between $\sigma^2 R$ and stellar mass in equally spaced bins of S\'ersic index; at a fixed stellar mass $\sigma^2 R$ and S\'ersic index are correlated. At fixed S\'ersic index, $\sigma^2 R$ appears to be directly proportional to stellar mass. The relations for different S\'ersic index bins are offset and nearly parallel. Figure \ref{fig:mdyn_sersic} strongly suggests that the dependence of $\sigma^2R$ on $D_n4000$ may be a consequence of the dependence of galaxy structure on $D_n4000$. }

\begin{figure}
\begin{center}
\includegraphics[width =  \columnwidth]{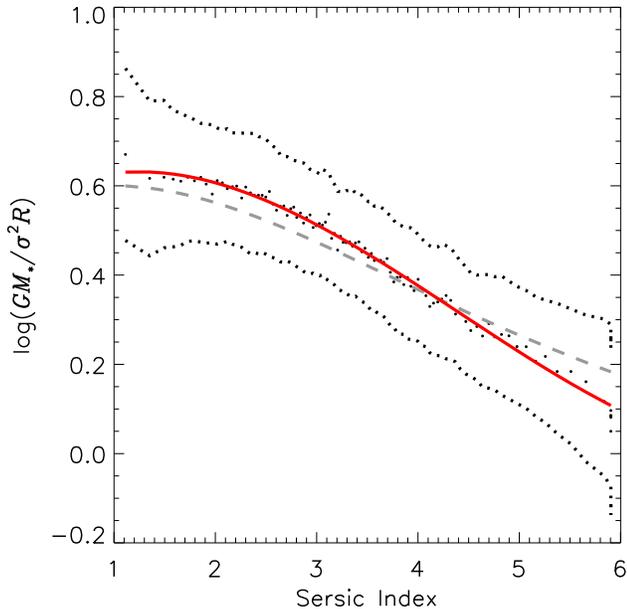}
\end{center}
\caption{Ratio of stellar mass to $\sigma^2R$, $K_\ast$, as a function of S\'ersic index. The black points are the median difference in 100 equally populated bins of stellar mass. The dotted lines are the limits of the central 50\% of the distribution. The red curve is the best-fit third degree polynomial to the black points and the gray dashed curve is the \citet{Bertin2002} relation given in Equation \ref{eq:k} but shifted by $-0.3$ dex. The \citeauthor{Bertin2002} relation is derived from the virial theorem and is independent of the data. Note the similarity between the red and dashed gray curve.}
\label{fig:sersic_res}
\end{figure}

The relation between the kinetic and potential energy of gravitational systems in virial equilibrium depends on the mass distribution, i.e. structure. Rearranging Equation \ref{eq:md} gives 
\begin{equation}
K_d = \frac{GM_d}{\sigma^2 R}.
\end{equation}
In analogy to $K_d$ we derive a quantity $K_\ast$ measured from the data by taking $M_\ast$ as a proxy for $M_d$, i.e. $K_\ast =GM_\ast/\sigma^2 R$. In Figure \ref{fig:sersic_res} we plot $K_\ast$ as a function of S\'ersic index. The quantity $K_\ast$ is indeed strongly correlated with the S\'ersic index. The best fit third degree polynomial (red curve in Figure \ref{fig:sersic_res}) is
\begin{equation}
K_\ast(n) = 0.55 +  0.14n - 0.066n^2 + 0.0050n^3.
\label{eq:kfit}
\end{equation}
Parameters of the $K_\ast(n)$ relation derived from the S\'ersic index measured in the $i$-band are consistent ($<1\sigma$) with Equation \ref{eq:kfit}. The relation between $K_\ast$ and $n$ is not a result of covariance between size and S\'ersic index measured in the same photometric band. 

We compare the $K_\ast(n)$ relation that we derive with the theoretical relation based on the virial theorem. The gray curve in Figure \ref{fig:sersic_res} shows the \citeauthor{Bertin2002} relation given in Equation \ref{eq:k} but shifted by $-0.3$ dex. The relative variation of $K_\ast(n)$ is consistent with the expectations of numerical calculations of $K_d(n)$ accounting for the dependence of the potential and kinetic energy on galactic structure.

$K_d$ relates the total dynamical mass to the velocity dispersion and size of galaxies. {The -0.3 dex offset between $K_\ast(n)$ and $K_d(n)$ occurs because the dynamical mass calculated using the \citeauthor{Bertin2002} relation is 0.3 dex larger than the stellar mass. Taken at face value this result either quantifies the dark matter contribution to the total dynamical mass or implies that an IMF heavier than Chabrier \citep[i.e.][]{Salpeter1955} is required.} However, the relative fraction of dark matter is degenerate with absolute uncertainties in the stellar mass and the offset is subject to systematic differences in the way $K_d$ is calculated relative to the way $K_\ast$ is measured. {We are unable to discriminate between a heavier IMF and/or non-negligible dark matter contribution.} The comparison of the relative variation, however, is robust. Henceforth we refer to the measured quantity $K_\ast(n) \sigma^2 R/G$ as the dynamical mass bearing in mind the absolute uncertainty in this quantity.

\begin{figure}
\begin{center}
\includegraphics[width =  \columnwidth]{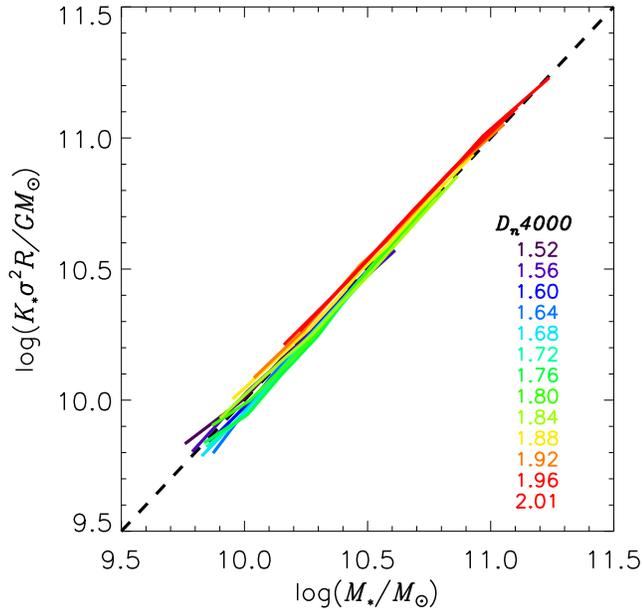}
\end{center}
\caption{Median $K_\ast \sigma^2 R/G$ as a function of stellar mass and $D_n4000$. $K_\ast$ is calculated using Equation \ref{eq:kfit}. Median $D_n4000$ values corresponding to the curves are listed and are color-coded to match the curves plotted. These values are the same as the median $D_n4000$ plotted as a function of stellar mass in Figure \ref{fig:vdisp_dn4000}B. The typical bootstrapped error for $\sigma^2R$ in each bin is $0.01$ dex. For clarity, error bars are omitted.}
\label{fig:kmdyn}
\end{figure}

The consistency between $K_d(n)$ and $K_\ast(n)$ confirms that the trend in Figure \ref{fig:sersic_res} can be explained by variations in galactic structure. Figure \ref{fig:kmdyn} shows dynamical mass as a function of the stellar mass and $D_n4000$. At a fixed stellar mass, the dynamical mass is virtually independent of $D_n4000$. This is true of the scatter as well (see Appendix). Thus, we conclude that the dependence of $\sigma^2 R$ on $D_n4000$ (Figure \ref{fig:mdyn_dn4000}) is primarily a consequence of the dependence of galaxy structure on $D_n4000$ (Figure \ref{fig:sersic_dn4000}).

Figure \ref{fig:kmdyn} shows that at a fixed stellar mass, there is a weak dependence of the dynamical mass on $D_n4000$. These trends may be related to metallicity variations, residual dependence on galaxy structure and/or correlations between $D_n4000$ and M/L ratio errors (among other possibilities). {We note that this dependence across the full range of $D_n4000$ is significantly weaker than the dependence apparent in Figure \ref{fig:mdyn_dn4000}.} 

\begin{figure}
\begin{center}
\includegraphics[width =  \columnwidth]{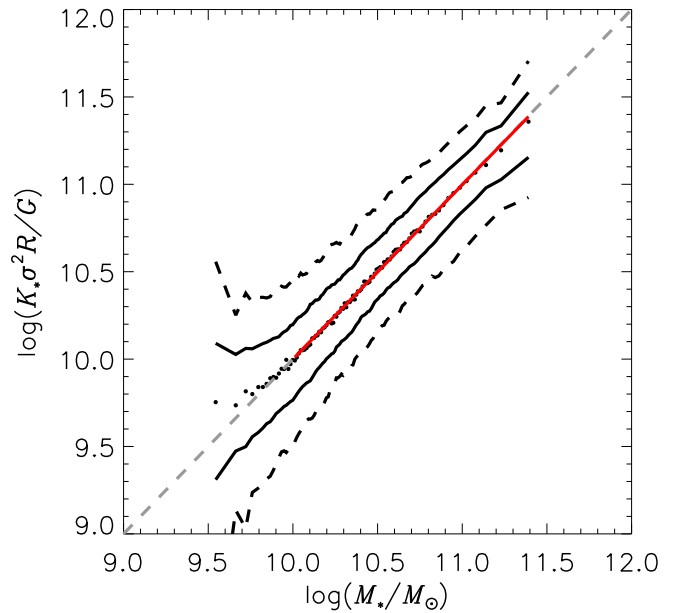}
\end{center}
\caption{Relative dynamical mass plotted as a function of stellar mass. $K_\ast$ is calculated from Equation \ref{eq:kfit}. Black points are median values of dynamical mass in 100 equally populated bins of stellar mass. Solid and dashed black curves show the limits containing the central 68 and 95\% of the distribution, respectively. The red curve is the best fit for $M_\ast>10^{10}M_\odot$. {The gray dashed line is the one-to-one relation. Only the slope of the relation is constrained by our procedure}.}
\label{fig:mass_mdyn}
\end{figure}

\begin{figure*}
\begin{center}
\includegraphics[width =  2\columnwidth]{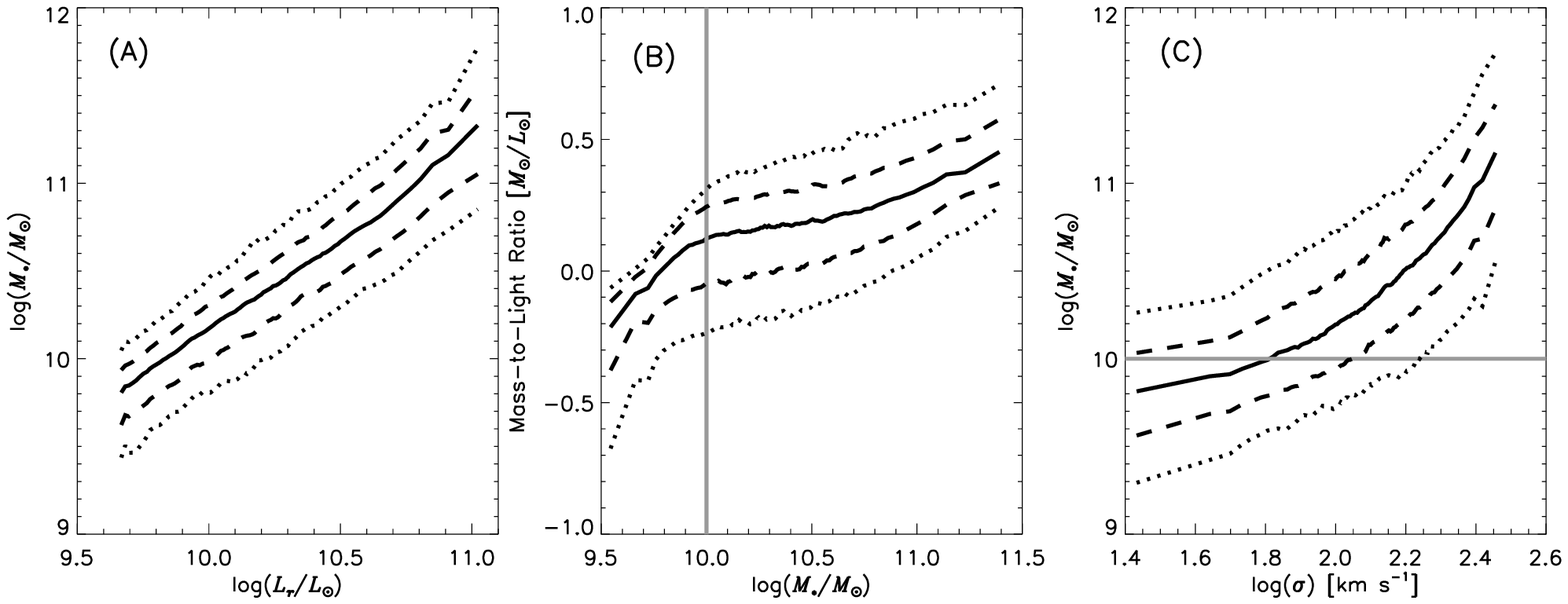}
\end{center}
\caption{(A) Stellar mass as a function of $r-$band luminosity. The solid curve is the median stellar mass in equally populated bins of luminosity. (B) The $r-$band stellar mass-to-light $(M_\ast/L_r)$ as a function of stellar mass. The solid curve is the median M/L ratio in equally populated bins of stellar mass. Note the incompleteness of the M/L ratio for galaxies with $M_\ast < 10^{10} M_\odot$; galaxies with the largest M/L ratios are missing. (C) Stellar mass as a function of velocity dispersion. The solid curve is the median stellar mass in equally populated bins of velocity dispersion. In each panel, the dashed and dotted curves indicate the intervals containing the central 68 and 95\% of the galaxy distribution, respectively. The gray lines in (B) and (C) indicate the stellar mass completeness limit of $M_\ast > 10^{10} M_\odot$. Note that only at log$(\sigma)\gtrsim2.3$ are 95\% of galaxies above the stellar mass completeness limit.}
\label{fig:completeness}
\end{figure*}

Figure \ref{fig:mass_mdyn} shows dynamical mass as a function of stellar mass. We fit the relation for galaxies with $M_\ast > 10^{10} M_\odot$ to avoid systematic bias due to selection effects\footnote{The slope we fit is insensitive to the stellar mass limit for $M_\ast \gtrsim 10^{9.7} M_\odot$}. We discuss this issue further in Section 5. The best-fit slope of the relation is $0.998 \pm 0.004$. Adopting the \citet{Bertin2002} relation to account for galaxy structure rather than our empirical relation, produces a slope of $1.008 \pm 0.004$. Applying no correction yields a slope of $1.084 \pm 0.005$. The slope of the relation between dynamical mass and stellar mass deviates significantly from unity when galaxies are assumed to be homologous systems. 

The direct proportionality between stellar mass and dynamical mass is robust to the method used to correct for galaxy structure. The \citet{Bertin2002} relation is completely independent of the data. Thus, it provides an important confirmation of our empirical correction.

We estimate the intrinsic $1\sigma$ scatter in the relation between stellar mass and dynamical mass by subtracting size and velocity dispersion measurement uncertainties in quadrature from the scatter of dynamical masses at a fixed stellar mass. For the velocity dispersion, we adopt the $1\sigma$ error value reported by the Portsmouth group. For size uncertainties, we adopt $\Delta$log$(R) = 0.11$ determined by comparing NYU group sizes to measurements based on Hubble Space Telescope imaging \citep[see Figure 1 in][]{Zahid2016a}. We measure an intrinsic scatter of $\sim0.2$ dex at a stellar mass of $10^{10} M_\odot$ and $\sim0.1$ dex at $10^{11} M_\odot$. These estimates are upper limits since they do not account for uncertainty in the stellar mass or $K_\ast(n)$ and because velocity dispersion errors may be underestimated by a factor of $\sim \sqrt{2}$ \citep{Fabricant2013, Zahid2016c}. Given that the observational uncertainties for the stellar masses alone are on the order of $\sim0.1$ dex, the intrinsic scatter in the relation between stellar mass and dynamical mass is significantly smaller than the 0.1 - 0.2 dex we quote.

\section{Stellar Mass as the Independent Variable}

\citet{Zahid2016c} demonstrate that the relation between stellar mass and velocity dispersion derived from magnitude limited surveys is biased when velocity dispersion is taken as the independent variable \citep[see also][]{Schechter1980}. Here, we analyze a volume limited sample to mitigate these selection biases. However, even for a volume limited sample, selecting stellar mass as the independent variable is necessary to minimize the impact of selection effects. The volume limited sample is limited in absolute magnitude and not velocity dispersion. Thus at any stellar mass the sample is fairly complete in velocity dispersion but the reverse is not true.

Figure \ref{fig:completeness}A shows the distribution of stellar mass as a function of the $r-$band luminosity. The M/L ratio has a significant spread at a fixed luminosity and both the ratio and scatter do not depend strongly on luminosity. The lack of luminosity dependence is a result of the $D_n4000$ selection \citep[see Figure 12 in][]{Geller2014}. At the luminosity limit, $L_r = 10^{9.66} L_\odot$, the highest mass galaxy has $M_\ast \sim 10^{10}M_\odot$. Thus, there is no stellar mass bias in the sample for galaxies with $M_\ast > 10^{10} M_\odot$. In other words, a galaxy with a stellar mass $>10^{10}M_\odot$ will be in the volume limited sample regardless of its M/L ratio. Below this mass limit, galaxies may be missing from the sample because of the spread in M/L ratios at a fixed luminosity.

Figure \ref{fig:completeness}B shows the M/L ratio as a function of stellar mass. At $M_\ast < 10^{10} M_\odot$, galaxies with high M/L ratios scatter out of the sample hence the sharp downturn in the distribution. At $M_\ast > 10^{10} M_\odot$ the M/L ratio distribution--and consequently the stellar mass distribution--is fully sampled. Thus, we fit the relation between stellar mass and dynamical mass (Figure \ref{fig:mass_mdyn}) for galaxies with $M_\ast > 10^{10} M_\odot$. The weak dependence of the M/L ratio on stellar mass for galaxies with $M_\ast > 10^{10} M_\odot$ occurs because massive galaxies tend to be older and thus have larger M/L ratios.


Figure \ref{fig:completeness}C shows the stellar mass distribution at a fixed velocity dispersion and highlights the problem with using the velocity dispersion as the independent variable. Figures \ref{fig:completeness}A and \ref{fig:completeness}B clearly demonstrate that at $M_\ast < 10^{10} M_\odot$, the stellar mass distribution is biased; galaxies with large M/L ratios are missing. Figure \ref{fig:completeness}C shows that the at a fixed velocity dispersion, the stellar mass distribution is extremely broad spanning $\sim1$ decade. Because of stellar mass incompleteness at $M_\ast<10^{10}M_\odot$, the velocity dispersion distribution is only unbiased at log$(\sigma)\gtrsim2.3$; only $\sim10\%$ of galaxies in our sample meet this criterion. The effect is particularly pernicious because the bias scales with the velocity dispersion, i.e. it is larger at smaller velocity dispersion. 

\begin{figure}
\begin{center}
\includegraphics[width =  \columnwidth]{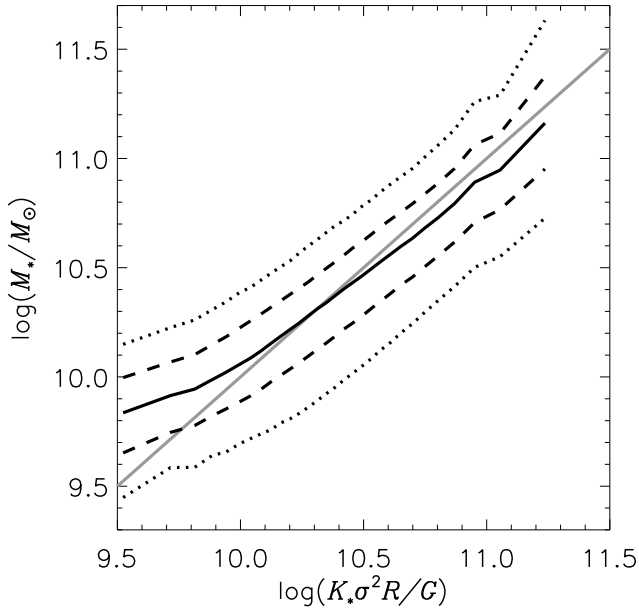}
\end{center}
\caption{Stellar mass as a function of the dynamical mass, $K_\ast \sigma^2 R/G$. $K_\ast$ is calculated from Equation \ref{eq:kfit}. The solid curve is the median stellar mass in equally populated bins of $K_\ast \sigma^2 R/G$. Dashed and dotted curves indicate the intervals containing the central 68 and 95\% of the galaxy distribution, respectively. Note the sub-linear slope of the relation which becomes progressively shallower at lower stellar masses because of incompleteness in the stellar mass distribution.}
\label{fig:slope}
\end{figure}

Figure \ref{fig:slope} shows stellar mass as a function of dynamical mass. Dynamical masses depend on the velocity dispersion and therefore the relation between dynamical mass and stellar mass is subject to the bias highlighted in Figure \ref{fig:completeness}. The relation has a slope $<1$ and progressively becomes shallower at lower stellar masses. This result is a direct consequence of the biased sampling of the full stellar mass range at a fixed velocity dispersion. At decreasing dynamical masses, a larger and larger fraction of the low stellar mass galaxies are missing. This stellar mass bias spuriously boosts the median stellar mass at lower velocity dispersion. In principle, one can model the bias and correct the relation in Figure \ref{fig:slope}. However, a more robust approach--and the one we use in our analysis--is to adopt stellar mass as the independent variable and fit the relation between stellar mass and dynamical mass only above the stellar mass completeness limit.

\section{Discussion}

The analysis of local quiescent galaxies presented here complements our previous efforts examining the dynamical properties of quiescent galaxies at intermediate redshift. Based on a highly complete spectroscopic sample, \citet{Zahid2016c} show that $\sigma \propto M_\ast ^{0.3}$ for quiescent galaxies with $z\lesssim0.7$. This scaling between stellar mass and velocity dispersion is the expected scaling for virialized systems \citep{Evrard2008}. The virial theorem prescribes a relationship between mass, velocity dispersion and size for galaxies. Thus, the results of \citet{Zahid2016c} are consistent with virial scaling but because accurate size measurements are not available for their sample, they do not uniquely constrain the virial equilibrium of galaxies. 

The fundamental plane is derived from luminosity (or stellar mass), velocity dispersion and size. The fundamental plane ostensibly reflects the approximate virial equilibrium of galaxies. \citet{Zahid2016a} show that the zero-point and orientation of the stellar mass fundamental plane does not evolve significantly for galaxies at  $z<0.6$. However, due to selection and systematic effects, \citet{Zahid2016a} only constrain the \emph{relative} variation in the zero-point and orientation. They are unable to measure an unbiased set of parameters for the stellar mass fundamental plane to compare with the expectations of virial equilibrium. 

A critical aspect of our investigation of the virial properties of quiescent galaxies is the combination of selection criteria and method of analysis. Our sample is derived from a spectroscopically complete ($\sim95\%$) parent sample. We mitigate selection effects by analyzing a volume limited sample taking stellar mass as the independent variable. We apply a very broad selection for quiescent galaxies based on the $D_n4000$ index. Our selection samples a large range in the observable properties relevant to the dynamics of quiescent galaxies, i.e. stellar mass, velocity dispersion, size, S\'ersic index. Moreover, we analyze our sample as a function of our selection parameter demonstrating that the results are insensitive to the $D_n4000$ selection threshold. This lack of sensitivity occurs because the relations in different bins of $D_n4000$ are nearly parallel.

The relation between stellar mass, velocity dispersion, size, S\'ersic index and $D_n4000$ has important implications for understanding quiescent galaxies. Here we discuss the implications for galaxy formation (Section 6.1) and quiescent galaxy evolution (Section 6.2). We also discuss the relation between stellar mass and dynamical mass (Section 6.3) and its implications for interpreting the fundamental plane and the approximate virial equilibrium of quiescent galaxies (Section 6.4).

\subsection{Theoretical Expectations}

At a fixed stellar mass, velocity dispersion and size depend strongly on the $D_n4000$ index. In the absence of active star formation, the $D_n4000$ index increases monotonically with time. Hence, older galaxies tend to be smaller and have larger velocity dispersions. Similar trends between velocity dispersion, size and age were found by \citet{Shankar2009} and \citet{vanderwel2009} based on ages determined from spectral indices \citep[see also][]{Forbes1999, Napolitano2010}. At a fixed stellar mass, velocity dispersion and $D_n4000$ are also anti-correlated in galaxy clusters \citep{Sohn2016}. These trends are generally expected in $\Lambda$CDM if galaxies form at different epochs.

\begin{figure*}
\begin{center}
\includegraphics[width =  2\columnwidth]{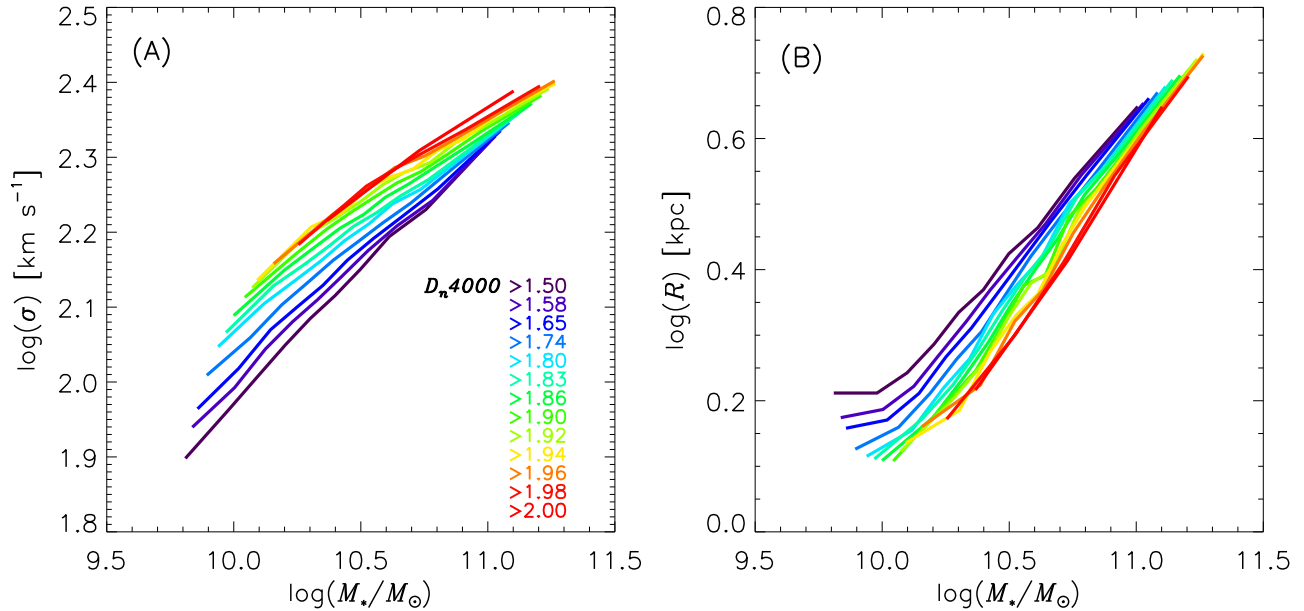}
\end{center}
\caption{(A) Median velocity dispersion in bins of stellar mass for the volume limited sample but with a progressively higher $D_n4000$ selection threshold. The $D_n4000$ selection in the legend is color matched to each curve. (B) Median size in bins of stellar mass based on the same $D_n4000$ selection cuts as in (A). }
\label{fig:dn4000_cut}
\end{figure*}

Dark matter decouples from cosmic expansion and collapses to form gravitationally bound structures if the density within some region exceeds the mean cosmic density by a fiducial factor \citep{Peebles1993}. The redshift dependence of the mean cosmic density is
\begin{equation}
\bar{\rho}_c \propto (1+z)^3.
\label{eq:rho_c}
\end{equation}
In a simple model where the collapse factor, $\bar{\rho}/\bar{\rho}_c$, is redshift independent, the mean density of collapsing material, $\bar{\rho}$, scales with the mean cosmic density. For a virialized collapsed halo,
\begin{equation}
M \propto r^3 \bar{\rho} \propto \sigma^2 r.
\label{eq:m_h}
\end{equation}
Here, $M, r$ and $\sigma$ are the mass, radius and velocity dispersion, respectively, and $M \propto \sigma^2 r$ follows from the virial theorem. Putting Equations \ref{eq:rho_c} and \ref{eq:m_h} together for a fixed halo mass
\begin{equation}
\frac{\sigma}{r} \propto (1+z)^{3/2}
\end{equation}
\citep[see also][]{Mao1998}. At a fixed mass, older galaxies should be smaller and have larger velocity dispersion. The observed relation between stellar mass, velocity dispersion, size and $D_n4000$ index are qualitatively consistent with this simple theoretical expectation, complex baryonic physics notwithstanding.

\subsection{Quiescent Galaxy Evolution}

At a fixed stellar mass, the S\'ersic index is strongly correlated with the $D_n4000$ index. A simple interpretation is that the correlation between S\'ersic index and $D_n4000$ reflects the redshift evolution of galaxy structure; galaxies that form earlier have larger S\'ersic indices. In this case, the S\'ersic indices of quiescent galaxies which cease star formation at early times should be larger. However, direct measurements report the opposite trend; high redshift quiescent galaxies typically have \emph{smaller} S\'ersic indices \citep{vanDokkum2010a, Buitrago2013}. Thus, direct measurements of high redshift galaxies appear to be inconsistent with the correlation between the S\'ersic index and $D_n4000$ measured for local quiescent galaxies. {This apparent disparity may be resolved if some physical process---for example, merging of galaxies---increases the S\'ersic index of quiescent galaxies after star formation ceases \citep[e.g.][]{Hilz2013}. }

We explore the characteristics of passive evolution by reconstructing the evolutionary history of quiescent galaxies based on our local sample. Here, by passive evolution we mean that the only change to a galaxy is the gradual aging of its stellar population. {This picture contrasts with an evolutionary scenario where quiescent galaxies continue to evolve via mergers even after they cease star formation.} Here, we \emph{assume} that galaxies evolve in a purely passive manner and infer the properties of the quiescent galaxy population at earlier epochs under this assumption. We test the assumption by comparing the inferred evolution we derive with direct measurements. 

In a purely passive evolutionary scenario, the quiescent galaxy populations of earlier epochs are fully represented in our sample and the dynamical and structural properties have not changed with time. We have defined quiescent galaxies as those with $D_n4000>1.5$ and the $D_n4000$ index of a quiescent galaxy increases monotonically as a function of time. In a purely passive evolutionary scenario, we can reconstruct the cosmic history of the quiescent galaxy population from the local quiescent galaxy population by applying a progressively higher $D_n4000$ selection threshold to our sample.

Figure \ref{fig:dn4000_cut}A shows the median velocity dispersion as a function of stellar mass for our sample with different minimum $D_n4000$s. At a fixed stellar mass, the median velocity dispersion increases as a function of the $D_n4000$ selection threshold. Figure \ref{fig:dn4000_cut}B shows the median size as a function of stellar mass for the same $D_n4000$ selection as in Figure \ref{fig:dn4000_cut}A. At a fixed stellar mass, the median size decreases as a function of the $D_n4000$ selection threshold. Figure \ref{fig:dn4000_cut} implies that in a purely passive evolutionary scenario for quiescent galaxies, the median velocity dispersion increases and the median size decreases as the redshift increases.  

\begin{figure*}
\begin{center}
\includegraphics[width =  2\columnwidth]{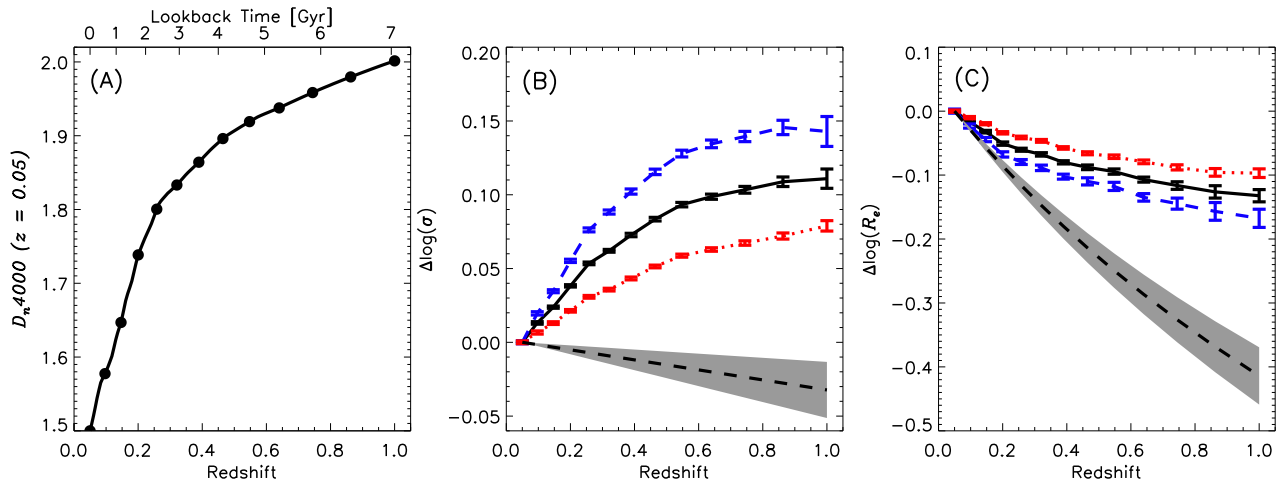}
\end{center}
\caption{(A) Quiescent evolution of the $D_n4000$ index projected back in time based on the FSPS model (see Section 2.5). The black curve relates the formation epoch of a quiescent galaxy to its $D_n4000$ index assuming purely passive evolution, i.e. the curve gives the redshift when a galaxy with a given $D_n4000$ index at $z=0$ had $D_n4000 = 1.5$. For example, all galaxies with $D_n4000 \gtrsim 1.85$ today would be classified as quiescent ($D_n4000 > 1.5$) at $z\sim0.4$. Conversely, local quiescent galaxies with $D_n4000 < 1.85$ have $D_n4000 < 1.5$ at $z>0.4$ and thus are not classified as quiescent at these earlier epochs. The black dots indicate the cuts applied to our sample in Figure \ref{fig:dn4000_cut}. (B) Relative zero-point offset of the median velocity dispersion of a galaxy with {$M_\ast = 10^{10.25}, 10^{10.5}$ and $10^{10.75} M_\odot$ (blue dashed, black solid and red dotted curves, respectively)} measured from fitting the median relation between velocity dispersion and stellar mass in Figure \ref{fig:dn4000_cut}A. The relative offset is plotted as a function of redshift using the relation between $D_n4000$ and redshift in (A). The dashed line is the zero-point of the median relation between stellar mass and velocity dispersion as a function of redshift from \citet{Zahid2016c} and the gray band shows the $2\sigma$ uncertainty. (C) Relative zero-point offset of the median size of a galaxy with {$M_\ast = 10^{10.25}, 10^{10.5}$ and $10^{10.75} M_\odot$ (blue dashed, black solid and red dotted curves, respectively)} measured from fitting the median relation between size and stellar mass in Figure \ref{fig:dn4000_cut}B. The relative offset is plotted as a function of redshift using the relation between $D_n4000$ and redshift in (A). The dashed line is the zero-point of the median relation between stellar mass and size as a function of redshift from \citet{vanderwel2014} and the gray band shows the $2\sigma$ uncertainty. {Note the differential inferred evolution of velocity dispersion and size as a function of stellar mass in (B) and (C), respectively.}}
\label{fig:dr_dm}
\end{figure*}

We test the passive evolutionary scenario by comparing the inferred evolution of velocity dispersion and size to direct measurements. Figure \ref{fig:dr_dm}A shows the passive evolution of $D_n4000$ measured from the FSPS model as a function of time. Using the model results, we can reconstruct the cosmic evolutionary history of the quiescent galaxy population by relating the $D_n4000$ selection threshold to redshift. For example, a galaxy with $D_n4000 \sim 1.85$ at $z\sim0$, has $D_n4000 >1.5$ at $z<0.4$ and thus would be classified as quiescent at $z<0.4$. Conversely, a local galaxy with $D_n4000<1.85$ would not be classified as quiescent at $z=0.4$. Thus, if galaxies evolve purely passively, a $D_n4000 > 1.85$ selection corresponds to the quiescent galaxy population at $z\sim0.4$. {Figure \ref{fig:dn4000_cut} shows that the passive evolution inferred from the correlation between $Dn4000$ selection threshold, velocity dispersion and size is mass dependent. We limit our analysis to a qualitative comparison of relative evolution of galaxies with $M_\ast = 10^{10.25}, 10^{10.5}$ and $10^{10.75} M_\odot$ and defer a detailed quantitative treatment to future efforts.} The conclusions are insensitive to the FSPS model parameters.

Figure \ref{fig:dr_dm}B shows the inferred redshift evolution of the median velocity dispersion assuming only passive evolution. We compare this result to direct measurements from \citet{Zahid2016c} who classify quiescent galaxies in the same way and find the zero-point evolves as $(1+z)^{-0.034 \pm 0.010}$. Figure \ref{fig:dr_dm}C shows the inferred size evolution compared to direct measurements from \citet{vanderwel2014} who find that early-type galaxy sizes evolve as $(1+z)^{-1.48 \pm 0.08}$. Other authors have reported evolution closer to $\sim(1+z)^{-1}$ \citep[e.g.,][]{Williams2010b}; our conclusions are the same if we compare to this weaker evolution in size. 

Direct measurements of velocity dispersion and size are inconsistent with the inferred evolution based on purely passive evolution of the quiescent galaxy population. The straightforward interpretation of these inconsistencies is that galaxies do not passively evolve after they cease star formation.

Disparities between the simple model and direct measurements in Figure \ref{fig:dr_dm} may be resolved if some physical process or processes increases the velocity dispersion, size and S\'ersic index of a quiescent galaxy after it ceases star formation. Several physical mechanisms have been proposed for the growth of the quiescent galaxy population. Our results rule out adiabatic growth \citep{Fan2008, Fan2010} as the primary mechanism because it requires that an increase in size is accompanied by a commensurate decrease in the velocity dispersion \citep[e.g.,][]{Hopkins2010}. Furthermore, the evolution of quiescent galaxy sizes can not \emph{solely} be a consequence of the fact that progenitors of the most recently formed quiescent galaxies are not part of the population at higher redshifts, the so-called progenitor bias \citep{vanDokkum2001, Carollo2013}. In other words, the redshift evolution of quiescent galaxy sizes can not be explained as a sole consequence of larger galaxies added to the population at late times. Our results are broadly consistent with growth driven by mergers \citep{Naab2009, Hopkins2010, Hilz2013}.

We qualitatively compare the archeological approach to galaxy evolution with direct measurements to conclude that quiescent galaxy evolution is not purely passive. We demonstrate this approach for galaxies at three stellar masses. {The inconsistencies between the inferred and directly measured evolution of size and velocity dispersion in Figure \ref{fig:dr_dm} appear to be mass dependent. Analysis of these inconsistencies will provide important quantitative constraints for understanding quiescent galaxy evolution.} A detailed application and analysis is beyond the scope of this paper. However, this exercise makes it clear that measurements of the kinematics and sizes of galaxies combined with a joint analysis of the number density evolution based on complete, consistently selected quiescent galaxy samples at several redshifts will provide an important test for various growth mechanisms proposed to explain quiescent galaxy evolution \citep[e.g., see][]{vanderwel2009}.

The rapid evolution of the $D_n4000$ index for $1.5<D_n4000<1.9$ makes it a very sensitive age proxy for newly formed quiescent galaxies. The $D_n4000$ index thus provides a robust means for studying quiescent galaxies. Figure \ref{fig:dr_dm} shows that the approach outlined here may constrain quiescent galaxy evolution based solely on samples of galaxies at modest redshifts ($z\lesssim0.5$) where all the necessary observations may soon be possible.

\subsection{Relation Between Stellar Mass and Dynamical Mass}

The stellar mass is directly proportional to the dynamical mass for galaxies in our sample. The simplest interpretation is that the stellar mass estimates are robust and that the dark matter contribution to the central dynamical mass is either a constant or negligible fraction of the stellar mass. The stellar mass estimates assume a constant IMF. Some studies suggest that the IMF \emph{systematically} varies in massive early-type galaxies \citep{vanDokkum2010b, Cappellari2012}. {Current dynamical and spectroscopic constraints on the IMF of individual galaxies appear to be inconsistent \citep{Smith2014} and thus systematic variations of the IMF remain uncertain. These inconsistencies may reflect the complicated character of IMF variations. For example, systematic variations of the IMF may be limited to the central regions of galaxies \citep[e.g.,][]{vanDokkum2016} and thus our total stellar mass estimates may be insensitive to such variations.} However, if the IMF \emph{systematically} varied globally, the apparent direct proportionality between stellar mass and dynamical mass would be a spurious coincidence. 

Several authors have examined the relation between stellar mass and dynamical mass for SDSS galaxies. \citet{Gallazzi2006} examine the relation making no correction for galaxy structure. They find that the power law index of the relation is $\sim0.8$ and the relation depends on the age of the stellar population. This dependence of the relation on stellar population age probably results from the dependence of galaxy structure on stellar population age (see Figure \ref{fig:sersic_dn4000}). If we take the dynamical mass as the independent variable and make no correction for galaxy structure, we recover a slope consistent with the \citet{Gallazzi2006} relation. In other words, if we analyze the data in a similar manner, our results and conclusions are consistent with \citet{Gallazzi2006}.  

\citet{Taylor2010b} examine the relation between stellar mass and dynamical mass but account for non-homology using the \citet{Bertin2002} relation. They also examine the stellar mass as a function of the dynamical mass and measure a power law index of 0.92 for the best-fit relation. However, \citet{Taylor2010b} also fit the relation between stellar mass and dynamical mass by minimizing the residuals in dynamical mass. In this case, they measure a power law index of unity. That is, if they take the stellar mass as the independent variable, they also find a direct proportionality between stellar mass and dynamical mass. When \citet{Taylor2010b} correct for non-homology and take the stellar mass as the independent variable, their scaling between stellar mass and dynamical mass is consistent with ours. This consistency is especially important because they use completely independent methods to measure sizes, galaxy profiles and velocity dispersions. {Thus, the results we find are robust to the analysis approach \citep[see also the appendix of][]{Taylor2010b}.}

The comparison of our results with \citet{Gallazzi2006} and \citet{Taylor2010b} highlights the importance of taking the stellar mass as the independent variable when examining relations involving the velocity dispersion. Failure to follow this approach leads to bias in the relation between stellar mass and dynamical mass (see Section 5 for detailed discussion).

\citet{Bolton2008} compare dynamical masses derived from the velocity dispersion and size to masses measured from strong lensing and find that that two are directly proportional. They conclude that quiescent galaxies in their sample form a homologous population and that no correction for galactic structure is required. Our results indicate that galactic structure does impact the dynamical properties of quiescent galaxies. If we do not apply corrections for non-homology, the relation we derive between dynamical mass and stellar mass is significantly steeper than unity, although it is consistent within the errors with the \citet{Bolton2008} relation. The strong lensing cross section scales as $\sim \sigma^4$ and the \citeauthor{Bolton2008} sample does not extend over a large range of S\`ersic indices given the selection effects \citep{Bolton2006, Bolton2008}. Figures \ref{fig:mdyn_dn4000} and \ref{fig:sersic_dn4000} imply that relations between stellar mass and $\sigma^2 R$ for a narrow range of S\`ersic indices are parallel and directly proportional \citep[see also][]{Taylor2010b}. Given selection effects, small sample size and the limited range of S\`ersic indices, the \citet{Bolton2008} sample may be insensitive to the impact of non-homology.

\subsection{The Fundamental Plane}

The fundamental plane is based on the virial properties of quiescent galaxies. However, it is well established that the parameters that define the fundamental plane are not the virial parameters. This deviation of the parameters is known as the ``tilt" of the fundamental plane. A number of solutions have been proposed. These include non-homology of the galaxy population, variation of the stellar populations, non-universal IMF and/or changing dark matter contribution or some combination of these effects \citep[to cite a few]{Djorgovski1995, Hjorth1995, Ciotti1996, Graham1997, Prugniel1997, Pahre1998, Scodeggio1998, Trujillo2004, Boylan-Kolchin2005, Jun2008, Allanson2009, Cappellari2006, Tortora2009, Bolton2008, Napolitano2010, Grillo2010b, Donofrio2013, Cappellari2013a}. No consensus on the origin of the tilt is yet established.

{Figure \ref{fig:mdyn_sersic} shows that the dynamical mass proxy, $\sigma^2 R$, and stellar mass are directly proportional at a fixed S\`ersic index. However, the relation between $\sigma^2 R$ and stellar mass at different S\`ersic indices are offset from one another. Galaxies are not uniformly distributed in the $\sigma^2 R$-stellar mass-S\`ersic index parameter space. Thus, deriving the relation between $\sigma^2 R$ and stellar mass without accounting for structural variations results in a relation which is tilted with respect to the the expectation from virial equilibrium.}

\citet{Cappellari2013a} examine the fundamental plane derived from integral field spectroscopy of 260 morphologically selected early-type galaxies. They derive the dynamical mass fundamental plane based on Jeans modeling and find that the mass fundamental plane is consistent with the predictions of virial equilibrium. The central dark matter fraction of galaxies in their sample is small and thus the dynamical mass is dominated by the stellar mass \citep[see also][]{Borriello2003}. They conclude that the tilt of the fundamental plane is entirely explained by systematic variations in the dynamical M/L ratios. We also conclude that the relation between stellar mass and dynamical mass implies that the central regions of quiescent galaxies are approximately in virial equilibrium. However, unlike \citet{Cappellari2013a}, our analysis indicates that tilt in the fundamental plane is partially due to non-homology. This difference may be a consequence of morphological sample selection and/or methodology used by \citet{Cappellari2013a}. 


The comparison of our results with \citet{Bolton2008} and \citet{Cappellari2013a} highlights the importance of a quiescent galaxy sample selection. Both \citet{Bolton2008} and \citet{Cappellari2013a} conclude that galaxies in their sample are homologous systems. However, \citet{Bolton2008} sample a limited range of S\'ersic indices which is likely a consequence of selection effects and \citet{Cappellari2013a} morphologically select their sample. Figures \ref{fig:mdyn_dn4000} and \ref{fig:sersic_dn4000} show that relation between stellar mass and $\sigma^2 R$ and S\'ersic index, respectively, are parallel at different $D_n4000$. The non-homology of quiescent galaxies in our sample is a direct consequence of our broad $D_n4000$ selection criteria for quiescent galaxies. If we make a more restrictive selection of quiescent galaxies based on the $D_n4000$ index, we would be insensitive to the effects of non-homology.

Our results suggest that stellar mass is directly proportional to the dynamical mass implying no tilt in the virial relation. Thus, the observed tilt in the fundamental plane may be a result of a combination of effects including stellar M/L ratio variations, non-homology and selection effects which may bias the relation. We account for M/L ratio variations by calculating the stellar mass and we account for non-homology by deriving an empirical correction based on the measured S\'ersic index. We mitigate selection bias by taking the stellar mass as the independent variable in the relation. Given that the fundamental plane is a multivariate relation and different techniques are used to fit the relation, it is not necessarily straightforward to account for all of these effects. 

\section{Summary and Conclusions}

We explore the relation between stellar mass and dynamical mass for a volume limited sample of quiescent galaxies with measured velocity dispersion, size, S\'ersic index and $D_n4000$. Use of these directly observable parameters leads to an unambiguous linear relation between stellar mass and a simple proxy for the dynamical mass. The relations among the observables provide a framework for probing and understanding the evolution of quiescent galaxies.

The relation between stellar mass, velocity dispersion and size depends strongly on $D_n4000$. At a fixed stellar mass, galaxies with large $D_n4000$ tend to have larger velocity dispersions and smaller sizes and vice versa. 

The relation between stellar mass and the dynamical mass estimator, $\sigma^2 R$, depends on $D_n4000$. The S\'ersic index of galaxies is also correlated with $D_n4000$. We empirically correct for non-homology using the S\'ersic index. The empirical correction is fully consistent with the analytical correction based on the virial theorem. The dependence of $\sigma^2R$ on $D_n4000$ appears to be a consequence of the lack of structural homology of the quiescent galaxy population. After correcting for non-homology, $\sigma^2R$ is directly proportional to the stellar mass suggesting that the dark matter contribution is either a constant or negligible fraction of the central dynamical mass. 

The $D_n4000$ index is a proxy for the age of the stellar population. Hence, our results indicate that galaxies forming at early times have larger velocity dispersions and smaller sizes. The standard $\Lambda$CDM formation scenario naturally leads to this dependence of velocity dispersion and size on galaxy age.

We test the assumption that quiescent galaxies evolve passively by using the measured $D_n4000$ index as a proxy of galaxy age at a fixed stellar mass. In this passive mode of evolution, the only change in the galaxy is the gradual aging of the stellar population. We vary the $D_n4000$ selection threshold to reconstruct the evolutionary history of quiescent galaxies. This reconstructed evolution is inconsistent with direct measurements. In other words, quiescent galaxies do not passively evolve.

Quiescent galaxies appear to be systems in approximate virial equilibrium. The direct proportionality between stellar mass and dynamical mass indicates that stellar masses calculated from broadband photometry and dynamical masses calculated from velocity dispersion and size are both robust mass estimators. The trends we observe are consistent with standard $\Lambda$CDM galaxy formation where quiescent galaxies do not simply passively evolve after they cease star formation.

The analysis based on stellar mass, velocity dispersion, size, S\'ersic index and $D_n4000$ at zero redshift provides a guide for a similar approach at greater redshifts. In fact, similar data covering the range of $z<0.5$ would provide surprisingly strong constraints on the evolution of quiescent galaxies. Imaging quiescent galaxies at intermediate redshifts (e.g., $z\sim0.5$) requires $\sim0''\!.1$ resolution to achieve a physical spatial resolution comparable to this study. This resolution can be achieved with space-based and/or adaptive optics observations. However, the spatial coverage of deeply imaged fields is limited to $<2$ deg$^2$ \citep[e.g., AEGIS, COSMOS, CANDELS;][]{Davis2007, Scoville2007, Grogin2011} and thus subject to cosmic variance. A combination of high quality medium resolution spectroscopy combined with deep high resolution imaging over large fields will provide robust quantitative constraints for determining the physical processes governing quiescent galaxy evolution. Future space based imaging surveys \citep[e.g., EUCLID][]{Laureijs2011} combined with large field-of-view multi-object spectrographs \citep[e.g., PFS;][]{Sugai2015} should transform our knowledge of quiescent galaxies.

\acknowledgements

HJZ gratefully acknowledges the generous support of the Clay Postdoctoral Fellowship. MJG is supported by the Smithsonian Institution. We thank Scott Tremaine and Charlie Conroy for stimulating and insightful discussion and Sirio Belli and Po-Feng Wu for constructive comments that improved the manuscript. We are grateful to Jubee Sohn for assistance with the data. We thank the anonymous reviewer for their careful reading of the manuscript and thoughtful comments. This research has made use of NASA's Astrophysics Data System Bibliographic Services. 

Funding for SDSS-III has been provided by the Alfred P. Sloan Foundation, the Participating Institutions, the National Science Foundation, and the U.S. Department of Energy Office of Science. The SDSS-III web site is http://www.sdss3.org/. SDSS-III is managed by the Astrophysical Research Consortium for the Participating Institutions of the SDSS-III Collaboration including the University of Arizona, the Brazilian Participation Group, Brookhaven National Laboratory, University of Cambridge, Carnegie Mellon University, University of Florida, the French Participation Group, the German Participation Group, Harvard University, the Instituto de Astrofisica de Canarias, the Michigan State/Notre Dame/JINA Participation Group, Johns Hopkins University, Lawrence Berkeley National Laboratory, Max Planck Institute for Astrophysics, Max Planck Institute for Extraterrestrial Physics, New Mexico State University, New York University, Ohio State University, Pennsylvania State University, University of Portsmouth, Princeton University, the Spanish Participation Group, University of Tokyo, University of Utah, Vanderbilt University, University of Virginia, University of Washington, and Yale University.

\appendix

\begin{figure*}
\begin{center}
\includegraphics[width =  \columnwidth]{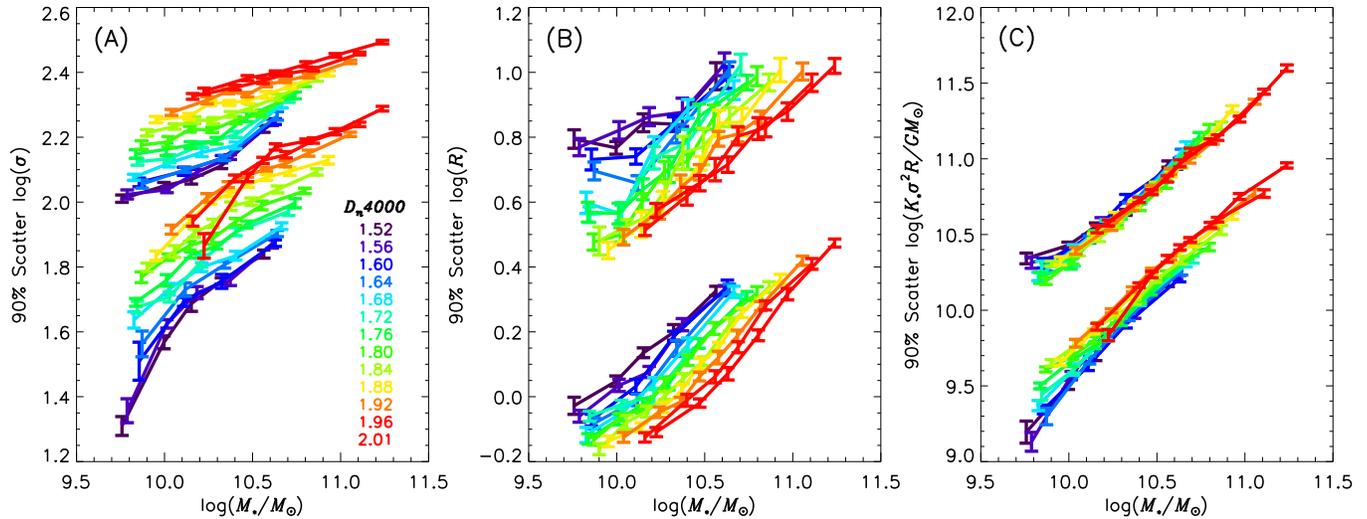}
\end{center}
\caption{Limits containing the central 90\% of the (A) velocity dispersion, (B) size (C) and dynamical mass distribution as a function of stellar mass and $D_n4000$. Error bars are bootstrapped. Median $D_n4000$ values corresponding to the curves are listed in (A) and are color-coded to match the curves plotted. These values are the same as the median $D_n4000$ plotted as a function of stellar mass in Figure \ref{fig:vdisp_dn4000}B.}
\label{fig:scatter}
\end{figure*}

We scrutinize the trends in Figures \ref{fig:vdisp_dn4000}, \ref{fig:size_dn4000} and \ref{fig:kmdyn} by examining the distribution of velocity dispersion and size as a function of stellar mass and $D_n4000$. Figures \ref{fig:scatter}A and \ref{fig:scatter}B show the limits containing the central 90\% of the velocity dispersion and size distribution at a fixed stellar mass, respectively. The trends in Figures \ref{fig:vdisp_dn4000} and \ref{fig:size_dn4000} result from a shift of the whole velocity dispersion and size distributions towards larger and smaller values at higher $D_n4000$, respectively; the median trends in Figures \ref{fig:vdisp_dn4000} and \ref{fig:size_dn4000} are representative of the entire distribution. Figure \ref{fig:scatter}C shows that after correcting for non-homologous galactic structure, the scatter of dynamical mass at a fixed stellar mass is nearly independent of $D_n4000$.

\bibliographystyle{aasjournal}
\bibliography{/Users/jabran/Documents/latex/metallicity}

 \end{document}